\documentclass[12pt,final]{iopart}
\usepackage{iopams}
\usepackage{amssymb}
\usepackage{xcolor}
\usepackage[titletoc]{appendix}
\renewcommand{\contentsname}{}
\usepackage{relsize}
\usepackage{cite}
\usepackage{bm}
\usepackage{array}
\newcolumntype{L}{>{$}l<{$}} 
\usepackage[
singlelinecheck=false 
]{caption}

\usepackage{xcolor}
\usepackage[titletoc]{appendix}
\usepackage{relsize}
\usepackage{bm}
\usepackage{braket} 
\usepackage{paracol}
\usepackage{graphicx}

\usepackage[displaymath, mathlines]{lineno}
\newcommand{\CQ}{{\mathcal Q}}

\newcommand{\CD}{\mathcal{D}}

\newcommand{\CR}{\mathcal{R}}

\newcommand{\CX}{{\mathcal X}}

\newcommand{\dd}{{\mathrm d}}
\newcommand{\pa}[1]{\left(#1\right)}

\newcommand{\say}[1]{`#1'}

\def\JCAP{{\it J.\ Cosmol.\ Astropart.\ Phys.}\ JCAP}

\def\PRD{{{Phys.\ Rev. D.}}}
\newcommand{\average}[1]{\left\langle #1 \right\rangle}

\newcommand{\initial}[1]{{#1_{\rm \bf i}}}

\usepackage{hyperref}
{
 \definecolor{BLACK}{gray}{0}
 \definecolor{WHITE}{gray}{1}
 \definecolor{RED}{rgb}{1,0,0}
 \definecolor{GREEN}{rgb}{0,1,0}
\definecolor{dgreen}{rgb}{.1,.6,.1}
\definecolor{BLUE}{rgb}{0,0,1}
 \definecolor{CYAN}{cmyk}{1,0,0,0}
 \definecolor{MAGENTA}{cmyk}{0,1,0,0}
 \definecolor{YELLOW}{cmyk}{0,0,1,0}
 \definecolor{aw}{rgb}{0.2,0.5,0.75}
  }
  \hypersetup{
    colorlinks,%
    citecolor=blue,%
    filecolor=blue,%
    linkcolor=magenta,%
    urlcolor=blue
}

\definecolor{MyB}{rgb}{0.1,0.1,1.0}

\definecolor{MyGreen}{rgb}{0.0,.5,0.0}

\definecolor{MyDarkRed}{rgb}{0.7,0,0}

\makeatletter
\long\def\@makefntext#1{\parindent 1em\noindent 
 \makebox[1em][l]{\footnotesize\rm$\m@th{^{\arabic{footnote}}}$}%
 \footnotesize\rm #1}
\def\@makefnmark{\hbox{$^{\arabic{footnote}}\m@th$}}
\def\@thefnmark{\arabic{footnote}}
\makeatother
\begin{document}
\def\beq{\begin{equation}} \def\eeq{\end{equation}}
\def\bea{\begin{eqnarray}} \def\eea{\end{eqnarray}}
\def \Scalar {S}
\def \Sc {A}
\def \VecField {\bm{V}}
\def \Vec {V}
\def \Bound {B}
\def \Vol {\mathcal{V}}
\def \norm {\mathcal{N}}
\def \heavy {\mathcal{H}}
\def \deltafun {\delta}
\newcommand\independent{\protect\mathpalette{\protect\independenT}{\perp}}
\def\independenT#1#2{\mathrel{\rlap{$#1#2$}\mkern2mu{#1#2}}}

\title[Solving the curvature and Hubble parameter inconsistencies]{Solving the curvature and Hubble parameter inconsistencies through structure formation-induced curvature}

\author{Asta Heinesen$^{*}$ and Thomas Buchert$^{*}$}
\address{$^*$Univ Lyon, Ens de Lyon, Univ Lyon1, CNRS, Centre de Recherche Astrophysique de Lyon UMR5574, F--69007, Lyon, France \\
\medskip
Emails: asta.heinesen@ens--lyon.fr and buchert@ens--lyon.fr}

\begin{abstract}
Recently it has been noted by Di Valentino, Melchiorri and Silk (2019) that the enhanced lensing signal relative to that expected in the spatially flat $\Lambda$CDM model poses a possible crisis for the Friedmann-Lema\^{\i}tre-Robertson-Walker (FLRW) class of models usually used to interpret cosmological data. 
The `crisis' amounts to inconsistencies between cosmological datasets arising when the FLRW curvature parameter $\Omega_{k0}$ is determined from the data rather than constrained to be zero a priori.  
Moreover, the already substantial discrepancy between the Hubble parameter as determined by Planck and local observations increases to the level of $5\sigma$.
While such inconsistencies might arise from systematic effects of astrophysical origin affecting the Planck Cosmic Microwave Background (CMB) power spectra at small angular scales, it is an option that the inconsistencies are due to the failure of the FLRW assumption. In this paper we recall how the FLRW curvature ansatz is expected to be violated for generic relativistic spacetimes.   
We explain how the FLRW conservation equation for volume-averaged spatial curvature is modified through structure formation, and we illustrate in a simple framework how the curvature tension in a FLRW spacetime can be resolved---and is even expected to occur---from the point of view of general relativity. 
Requiring early-time convergence towards a Friedmannian model with a spatial curvature parameter $\Omega_{k0}$ equal to that preferred from the Planck power spectra resolves the Hubble tension within our dark energy-free 
model.
\end{abstract}
\setcounter{footnote}{0}
\section{Introduction} \label{sec::intro}
Since the founding of relativistic cosmology, the FLRW 
class of models has been used to interpret cosmological data and to constrain the dynamical nature of our Universe. While the FLRW 
spacetimes offer a simple framework for interpreting cosmological data, the spacetimes which the FLRW models can reasonably approximate are limited. 
In particular, the FLRW curvature ansatz of a single constant parameter describing the curvature of space throughout the evolution of the Universe excludes the general-relativistic coupling\footnote{In this paper, the word \say{coupling} refers to the interaction of geometry and matter given by Einstein's field equations, and should not be confused with any additional coupling introduced in modified gravity scenarios.} of spatial curvature to the matter sources. Such dynamical coupling is in general expected to be non-cancelling even when averaged over the largest scales. 

It is worth recalling these limitations of the FLRW ansatz given the inconsistencies in cosmological parameters inferred by various experiments when using the FLRW framework for data reduction\cite{Valentino,Valentino2,Handley,Riess2018,Riess2019,kids}. 
While the inconsistencies highlighted in \cite{Valentino,Valentino2,Handley} might partly be due to unknown astrophysical phenomena affecting the high multipoles of the Planck power spectra \cite{Planck2015,Planck2018}, systematics in supernovae data reduction \cite{RameezHubble}, and the statistical methods used \cite{Efstathiou}, we believe that there is reason to consider the possibility that the inferred discrepancies in cosmological parameters between datasets could be the result of neglected physics in the FLRW class of models.   
Various phenomenological extensions of the `base' $\Lambda$CDM model (Cold Dark Matter and a cosmological constant $\Lambda$) with six parameters have been investigated for their potential to solve the parameter discrepancies, including for instance non-minimal dark sector physics and running of the spectral index \cite{ValentinoNonminimal,ValentinoExtended}. However, the effect of changing parameters within the FLRW paradigm appears to fall short with respect to the significance of the Hubble tension \cite{Riess2020}. We argue that the physics driving the tensions between datasets within the $\Lambda$CDM paradigm might simply be general-relativistic interaction between structure in the matter distribution and curvature which generically introduces extra terms on the largest scales to the Friedmann equations of a strictly structureless universe model. 
Spatially averaging the Einstein field equations introduces non-cancelling correction terms to the large-scale evolution equations of FLRW model spacetimes. For example, on a compact domain $\CD$, the volume-averaged variance of the expansion rate $\Theta$, $\average{(\Theta - \average{ \Theta}_\CD)^2}_\CD$, acts as source of an effective Hubble rate $H_\CD = 1/3 \average{\Theta}_\CD$,  that positively accumulates differences in expansion rates, say between that of clusters and voids, from the smallest up to the largest scales. This variance counteracts gravity and couples to the average scalar curvature, and thus modifies the average dynamics relative to that expected in FLRW universe models.
Hence, the aforementioned tensions might be solved from first principles within general relativity (GR) without the need for introducing dark energy or for introducing phenomenological parameters or exotic physics. 

In this paper we highlight the differences between curvature in FLRW cosmology\footnote{When we refer to \say{FLRW} in this paper, we shall mean the class of general relativistic FLRW solutions, though some of the statements made would generalize to modified gravity scenarios.} and in generic relativistic spacetimes, focusing on spacetimes with a single irrotational dust source. 
We illustrate how dynamical curvature expected from general relativity might account for the tensions encountered in the FLRW framework. We invoke---as a show-case and proof of concept---a simple and physically motivated solution to an exact scalar averaging scheme that quantifies spacetime dynamics on the largest scales. 
We emphasize that an FLRW solution for interpreting cosmological data can only make physical sense if it describes the Universe on average.

We start by discussing properties of spatial curvature, relations to topology and conservation laws in section \ref{sec:curvature}. We introduce the spatially averaged Einstein equations as formulated in the Buchert scheme in section \ref{sec:avfield}. Then, we present a class of models that respects generic dynamical properties of average spatial curvature in a general-relativistic spacetime, and we employ a dark energy-free model that solves the curvature and Hubble parameter inconsistencies in section \ref{sec:scaling}. We discuss our results in relation to simulation studies in section \ref{sec:discussion}. We present our conclusions in section \ref{sec:conclusion}.

\section{Remarks on spatial curvature}
\label{sec:curvature}

In general relativity, information about the curvature of spacetime is fully contained in the Riemann tensor. In FLRW cosmology the existence of six killing vector fields---which represent translational and rotational invariance and are physically motivated by the large-scale statistical homogeneity and isotropy suggested by cosmological data---is assumed in order to reduce the space of metric solutions.
In the FLRW class of spacetimes the Riemann tensor of spatial hypersurfaces is completely determined by the three-dimensional Ricci scalar $\mathcal{R} = 6 k/a^2(t)$, where $k$ is the constant-curvature parameter of dimensions $1/\textrm{length}^2$, and $a(t)$ is the dimensionless scale factor evaluated at the hypersurface labelled by the time parameter $t$. 
The curvature parameter $k$ can be understood as an integration constant in a Newtonian derivation of the Friedmann equations, representing the conserved energy of accelerated particles located on the edge of an isolated uniformly expanding sphere. 

Assuming simply-connected three-manifolds, the sign of $k$ determines the topology of the spatial sections, such that $k>0$ implies the topology of a hypersphere, $k=0$ implies Euclidean topology, and $k<0$ implies hyperbolic space. 
For generic spacetimes there are no `quantized' scalar curvature states describing the topology of space. 
These topological implications hold because the scalar curvature coincides with the sectional curvatures of the manifold in this highly symmetric case. 
Contrary to what is the case for the FLRW class of spacetimes, the Universe may be described by spherical topology on spatial hypersurfaces while being equipped with a metric that everywhere has \textit{negative} three-dimensional Ricci scalar curvature over the same hypersurfaces \cite{Lohkamp}. 
\textit{It does therefore not in general make physical sense to draw conclusions on the topology of the Universe based on the three-dimensional Ricci scalar}. 
Moreover, several studies of inhomogeneous cosmological models point towards average negative three-dimensional Ricci scalar as an attractor in the late Universe irrespective of the exact initial conditions given at the CMB epoch. (See, e.g., \cite{buchertcarfora:curvature,roy:instability,buchertrasanen,Rasanenreview,Bolejko}.) This generic feature is physically explained by (almost) empty void regions gaining volume dominance in the late Universe.   

The FLRW three-dimensional Ricci scalar is associated with the conservation law $\mathcal{R} a^2(t) = const.$
In general, there does not exist such an integral constraint for the volume-averaged three-dimensional Ricci scalar as in the FLRW class of models. That is, the average scalar curvature does not obey a conservation law like the average 
restmass density \cite{buchertcarfora:curvature}. It turns out, however, that there exists an integral constraint that couples the volume-averaged scalar curvature to the structure inhomogeneities, generalizing the FLRW conservation equation \cite{dust,perfectfluid,generalfluid}. For a spacetime with a single irrotational dust source with four-velocity $\bm u$ the integral constraint as formulated within the scalar averaging scheme \footnote{See the following section for more introduction.} reads \cite{dust}:
\begin{equation} \label{eq:integrability}
\frac{1}{a_\mathcal{D}^6} \, (\, \mathcal{Q}_\mathcal{D} \,\;a_{\mathcal{D}}^6 \,)^{\bdot} %
\,+\, \frac{1}{a_\mathcal{D}^{2}} \,(\,\braket{\mathcal{R}}_\mathcal{D} a_\mathcal{D}^2 \,)^{\bdot}\, \;=\;0\ ,
\end{equation}
where $\mathcal{R}$ is the three-dimensional Ricci scalar defined on the spatial hypersurfaces normal to the fluid four-velocity.   
The averaging operation $\braket{.}_\mathcal{D}$ is the Riemannian average over a subdomain $\mathcal{D}$ of the same set of hypersurfaces, and ${}^{\bdot} \equiv {\mathrm d}/{\mathrm d}\tau$ denotes the derivative with respect to the proper time function $\tau$ of the fluid.\footnote{The proper time function $\tau$ is defined uniquely from the family of possible proper time functions by requiring $\bm u = \bm \nabla \tau$.}   
The spatial domain $\mathcal{D}$ is defined to follow the fluid flow (no net flow of fluid elements into and out of of the averaging domain) but might otherwise be chosen for the physical problem at hand. The volume of the domain, normalized by the initial volume, defines an effective dimensionless scale factor on the domain: $|\CD| / |\initial{\CD}| = : a_\CD^3$.

The function $\mathcal{Q}_\mathcal{D}$ is the `kinematical backreaction',\footnote{For a detailed discussion on kinematical backreaction see \cite{dust,Buchert2008}.} and is defined from the variance of the rate of expansion and the averaged shear scalar of the fluid congruence over the domain $\mathcal{D}$, ${\mathcal Q}_\CD \equiv \frac{2}{3} \average{(\Theta - \average{\Theta}_\CD)^2} - 2 \average{\sigma^2}_\CD$. 
The average spatial curvature is generically not separately preserved but couples to the spacetime structure through $\mathcal{Q}_\mathcal{D}$. This in turn gives rise to an effective dark energy-like effect \cite{dust}. A coupling of this type is expected from first principles, since curvature generically couples to structure in the matter distribution. The dependence on domain of the conservation equation reflects the regional departure from homogeneity and isotropy which is in general present in a universe model with structure.  
For a structureless and isotropically expanding fluid $\mathcal{Q}_\mathcal{D} = 0$ and \eref{eq:integrability} reduces to the FLRW conservation equation for $\mathcal{R}$ for all domains $\mathcal{D}$. The no-backreaction conservation equation $(\braket{\mathcal{R}}_\mathcal{D} a_\mathcal{D}^2 )^{\bdot} = 0$ remains true to linear order in perturbation theory around a FLRW background.\footnote{\label{Qzero}$\mathcal{Q}_\mathcal{D}$ also vanishes if deviations from homogeneity are evaluated on flat space sections with periodic boundary conditions in Newtonian cosmology\cite{buchertehlers,buchert:newton}. This result carries over to some relativistic perturbative settings, where the background of the perturbations is assumed to be spatially flat. It has for instance been shown that backreaction in a dust universe model can be formulated as a boundary term even up to second order in standard perturbation theory in the comoving synchronous gauge \cite{NanLi}. In first order post-Newtonian theory with averaging performed over hypersurfaces in a Poisson gauge-adapted foliation, deviations of the averaged Einstein equations from the Friedmann equations have been shown to reduce to a boundary term as well \cite{Clifton}. However, this result presupposes the average density function of the domain to coincide with the background density of the  FLRW background solution, a property which is in general not satisfied in inhomogeneous spacetimes.}
The integral constraint \eref{eq:integrability} suggests that the FLRW class of solutions forms a measure zero set (no interaction with structure) and that fine-tuning or restricting assumptions leading to exact cancelation are in general required to maintain a notion of conservation of curvature during a time interval and at some given spatial scale. In particular, it is expected for inhomogeneous universe models that the FLRW curvature conservation equation is violated at the onset of structure formation \cite{buchertrasanen,Rasanenreview}. 
Furthermore, the exactly zero-curvature FLRW model forms a measure zero set within the FLRW class of models. 
The FLRW models have been shown to be globally gravitationally unstable in the directions of the dark energy and dark matter sectors, i.e. the average model is driven away from the FLRW solution, which forms a repeller within a dynamical systems analysis in the cases where $\mathcal{Q}_\mathcal{D}$ mimics the dark components \cite{roy:instability}. 
We also note that it is a generic feature of relativistic spacetimes that average spatial curvature $\braket{\mathcal{R}}_\mathcal{D}$ can change sign over cosmic epochs which is impossible in the FLRW class of models.

Table \ref{tab:summarycurvature} shows a summary of important properties related to spatial curvature in the FLRW class of spacetimes and how these properties generalize within full GR. 
The FLRW class of GR spacetimes are contained in the full GR case, but constitute a measure zero set within the full set of GR solutions.

\begin{table}[h]\footnotesize

\caption{Comparison of curvature properties within the FLRW class of cosmological models and for generic averaged globally hyperbolic spacetime models.}
\begin{tabular}{| l | l | l | }
\hline
  & FLRW  & Average within generic GR \\
  \hline
\parbox[t]{3.2cm}{Topology}  & \parbox[t]{5.6cm}{sign$(\mathcal{R})$ determines the spatial topology for simply-connected domains\vspace{5pt}} & \parbox[t]{5.5cm}{$\braket{\mathcal{R}}_\mathcal{D}$ does not in general allow conclusions on topological properties}   
\\ \hline
\parbox[t]{3.2cm}{Integral constraint}  &\parbox[t]{5.6cm}{local `Newtonian' energy conservation: $(\,\mathcal{R} a^2 \,)^{\bdot} = 0$ } & \parbox[t]{5.5cm}{general-relativistic coupling of $\braket{\mathcal{R}}_\mathcal{D} $ to structure: \\ $\frac{1}{a_\mathcal{D}^6} \, (\, \mathcal{Q}_\mathcal{D} \,\;a_{\mathcal{D}}^6 \,)^{\bdot} +  \frac{1}{a_\mathcal{D}^{2}} \,(\,\braket{\mathcal{R}}_\mathcal{D} a_\mathcal{D}^2 \,)^{\bdot} = 0 \; $ \vspace{5pt}}  \\
  \hline
\parbox[t]{3.2cm}{Sign of curvature}  & \parbox[t]{5.6cm}{sign$(\mathcal{R})$ is preserved throughout the evolution of the Universe and on all scales\vspace{5pt}} & \parbox[t]{5.5cm}{sign$(\braket{\mathcal{R}}_\mathcal{D})$ can change in response to structure in the spacetime and may vary on different scales }  \\
\hline  
\parbox[t]{3.2cm}{Copernican principle}  & \parbox[t]{5.6cm}{satisfied in its most strict interpretation. All fundamental observers are subject to the same local curvature\vspace{5pt}} & \parbox[t]{5.5cm}{can be satisfied in a weaker sense than for FLRW. `Distributional equivalence' between observers}   \\
\hline  
  \end{tabular}
\label{tab:summarycurvature}
\end{table}
\vspace{-3pt}
The complexities introduced when considering full GR, which is \textit{a priori} a background-free theory, carries over to perturbative settings. In FLRW-based perturbative frameworks physical geometric and matter fields are defined with respect to an FLRW background 
spacetime, relative to which they must be assumed to be small (and of similar order of magnitude). 
A generic spacetime is of course not restricted by such smallness assumptions relative to a global background.    

When there is not necessarily a global spacetime solution obeying exact symmetries constituting a background of all cosmological matter fields, perturbation theory becomes hard to handle and even ill-defined. 
Examples of difficulties are the identification of a good background spacetime (as the average over inhomogeneities), the interpretation and uniqueness in definitions of the \say{fields} living on the background, and the break-down of standard Fourier analysis when the identified background is curved. Perturbations are often treated as if they propagate on a flat background spacetime, while in reality perturbations propagate in the perturbed spacetime. For such intrinsic perturbation schemes in comparison with standard perturbation schemes the reader is directed to \cite{rza4} and \cite[sect.7.4.2]{Buchert2011}).

\smallskip

One might argue that the FLRW solutions are compatible with the high degree of isotropy of the CMB together with the Copernican principle, and that the FLRW constraints on spatial curvature are natural in a physical universe model with no preferred observers. 
While it is true that comoving observers in the FLRW spacetimes can be considered strictly identical, 
the FLRW models constitute an idealized limit of realistic statistically homogeneous and isotropic models. 
In 1968 Ehlers, Geren, and Sachs proved that for a solution of the Einstein equations with the only matter source being a radiative fluid with an isotropic distribution function, the spacetime is either stationary, or given by an FLRW solution, or a special solution with non-zero rotation and acceleration of the radiation fluid \cite{EGS}.
These results have been generalized to the case of a radiative fluid with an \textit{almost} isotropic distribution function along with realistic matter content \cite{almostEGS} as is relevant for our observations of the CMB. Here it was shown that if the CMB temperature and its derivatives are almost isotropic everywhere in a dust-dominated and expanding universe model, and the observers of the CMB are geodesic, the spacetime is \textit{almost} described by an FLRW metric. 
As pointed out in \cite{EGSRasanen}, the results in \cite{almostEGS} follow from the assumptions about smallness of derivatives in the temperature field of the CMB photons, which are not directly observed and which are directly related to local derivatives of the metric tensor which are expected to be large in the real Universe.
However, as a conservative assumption, we may adopt an \say{almost}-FLRW model in the radiation-dominated phase and an ample time thereafter. We expect any significant deviations from the FLRW class of models to emerge at the time of onset of structure formation. 

\section{Averaging of the Einstein equations}
\label{sec:avfield}
Here we provide a brief introduction to the scalar averaging scheme (the `Buchert equations'). For an overview of this averaging scheme see, e.g., \cite{dust,perfectfluid,generalfluid,Buchert2008,buchertrasanen,ellisbuchert}.
The Buchert scheme of inhomogeneous cosmology replaces local spacetime variables by 
volume-averaged variables which represent the `macro-state' on a given domain of the spacetime. The global dynamics is constrained by the local spacetime variables which obey the Einstein equations. Consequently the macroscopic variables must obey a set of equations which are similar in form to the Friedmann equations, but with additional terms accounting for the large-scale integrated effect of local inhomogeneity and anisotropy, named `cosmological backreaction'. 
A strength of this averaging scheme is that it offers a consistent framework for analyzing large-scale properties of a spacetime through a few global variables, without necessarily having full knowledge of metrical properties of the spacetime on smaller scales. It offers a complementary framework to FLRW cosmology for studying cosmology at the largest scales, and for making explicit the assumptions which must in practice be made for a spacetime solution to provide a reasonable approximation for the average dynamics.

We consider Einstein's equations $G_{\mu \nu} + \Lambda g_{\mu \nu} = 8\pi G T_{\mu \nu}$ for an irrotational dust universe model, where the matter is comprised of an irrotational and geodesic congruence of world-lines with four velocity $u^{\mu}$ and density $\varrho$ such that $T_{\mu \nu}=\varrho u_{\mu } u_{\nu}$. 
By averaging projections of the Einstein equations in the fluid frame we can arrive at a set of evolution equations of cosmologically relevant macroscopic variables in fluid proper time. For a detailed derivation of the below equations, see \cite{dust}. 

Averaging the local Raychaudhuri equation in the fluid rest frame over a spatial domain $\CD$ comoving with the fluid, we arrive at the \emph{averaged} Raychaudhuri equation,
\begin{equation}
\label{eq:averagedraychaudhuri}
3\,\frac{{\ddot a}_\CD}{a_\CD} \,+\, 4\pi G \, \average{\varrho}_{\CD} \,-\,\Lambda\;=\; {\CQ}_\CD\ ,
\end{equation}
where $a_\CD \equiv ( |\CD| / |{\initial\CD}|)^{1/3}$ is the `volume scale factor' as introduced above, $\average{.}_{\CD}$ is the covariant averaging in the fluid frame\footnote{The dependence on foliation is expected to be weak on large scales in physical applications of this covariant averaging scheme. This has been discussed  in \cite{BMR} and contrasted with coordinate-dependent statements in the literature. We note that the choice of comoving foliation applied here can be defined 
coordinate-independently and is distinct from the choice of gauge in standard model perturbation theory \cite{Clifton}. 
For the explicit demonstration of $4-$covariance of the averaging formalism adapted in this paper, see \cite{covariance}.} over the comoving spatial domain $\CD$, $\Lambda$ is the cosmological constant, and the overdot denotes the covariant time-derivative, $\dot{} \equiv \frac{\dd}{\dd t}$, where $t$ is the proper time of the fluid which defines level hypersurfaces orthogonal to the fluid flow. 
$\CQ_\CD$ is the \say{kinematical backreaction},
\begin{equation}
\label{eq:backreactiondef}
{\mathcal Q}_\CD \equiv \frac{2}{3} \average{(\Theta - \average{\Theta}_\CD)^2} - 2 \average{\sigma^2}_\CD \, . 
\end{equation}
The averaged Raychaudhuri equation \eref{eq:averagedraychaudhuri} is analogue to the acceleration equation of FLRW cosmology, but has the kinematical backreaction as an additional source term.  
The kinematical backreaction variable is composed of two non-negative functions---one proportional to the variance of the isotropic expansion rate $\theta$, and one involving the average of the squared rate of shear $\sigma^2$---which will not be zero individually on any scale for non-trivial spacetimes. Vanishing of backreaction thus requires exact balance between the two terms or, globally, a reduction of ${\mathcal Q}_\CD$ to boundary terms together with the assumption of a boundary-free space form.

This is also true in universe models with a notion of statistical homogeneity and isotropy. Such models are expected to approach a \say{monopole} state on the largest scales, but the Friedmannian monopole state will in general acquire correction terms in the form of backreaction from the structure on smaller scales. 

When the spatial variance of the isotropic expansion scalar dominates over the average of the squared shear scalar, the kinematical backreaction term $\mathcal{Q}_{\mathcal D}$ is positive and acts as a driver for volume average acceleration of space. 
This is expected to happen at large spatial scales due to the growing difference between expansion rates in voids and virialized objects, while shear is dominant on smaller scales where anisotropic structures are observed. 
The shear scalar is not individually constrained in the analysis of this paper, as there is a trade-off between variance of the isotropic expansion scalar and the shear scalar in the kinematical backreaction variable (which is the fundamental variable of this analysis). 
However, since we consider physics on the largest scales in this paper we expect no `net shearing effect'. 
The averaged inhomogeneous models offer the possibility to analyze data on smaller scales as well, and in turn determine the role of regional shear in the models.

The local Hamiltonian constraint equation can be averaged in a similar way resulting in the \emph{averaged} Hamiltonian constraint equation,
\begin{equation}
\label{eq:averagedhamilton}
3\left( \frac{{\dot a}_\CD}{a_\CD}\right)^2 \,-\, 8\pi G \,\average{\varrho}_{\CD}\,-\,\Lambda \;=\; - \, \frac{\average{\CR}_{\CD}\,+\,{\CQ}_\CD }{2}\,,
\end{equation}
where $\average{\CR}_{\CD}$ is the average of the spatial three-Ricci scalar in the fluid frame. 
Interpreting the kinematical backreaction as a fluid component, a positive kinematical backreaction term contributes with negative energy density and pressure and acts as a source of average volume acceleration. 
Positive volume acceleration can thus emerge globally without introduction of exotic matter components violating energy conditions. In physical scenarios positive backreaction emerges on large scales due to the spatial variance in expansion rate (e.g. between voids and overdense regions), and is accompanied by growing volume average negative curvature. The large-scale notion of accelerating space is in such scenarios to be interpreted simply as faster expanding regions taking over more volume relative to slowly expanding regions, and through this contributing more to the average expansion rate and spatial curvature. 

We define the domain-dependent volume Hubble rate $H_\CD \equiv \dot{a}_\CD/a_\CD$, and write the averaged Hamiltonian constraint equation on the form 
\begin{equation}
\label{eq:hamiltonianC}
\Omega_{m}^{\CD} +\, \Omega_\Lambda^\CD \,+\, \Omega_{\CR}^{\CD} \,+\, \Omega_{\CQ}^{\CD} \;=\, 1   \, , 
\end{equation}
where the four domain-dependent cosmological `parameters' $\Omega_m^\CD$, $\Omega_\Lambda^\CD$, $\Omega_\CR^\CD$, and $\Omega_\CQ^\CD$ constituting a `cosmic quartet' are defined by:
\begin{equation}
\label{eq:omega}
\fl
\Omega_{m}^{\CD} \;\equiv\; \displaystyle{\frac{8\pi G}{3 H_{\CD}^2} \, \average{\varrho}_{\CD}}\ ; \qquad \Omega_{\Lambda}^{\CD} \;\equiv\; \;  \frac{\Lambda}{3 H_{\CD}^2} \ ; \qquad
\Omega_\CR^\CD \;\equiv\;  - \, \displaystyle{\frac{\average{\CR}_{\CD}}{6 H_\CD^2}} \ ; \qquad \Omega_{\CQ}^{\CD} 
\;\equiv\;  - \,\frac{{\CQ}_{\CD}}{6 H_{\CD}^2 }\, .
\end{equation}
The average of the local energy-momentum conservation equation reads:
\begin{equation}
\label{eq:energyconservation}
\average{\varrho}_{\CD}^{\bdot} \,+\, 3\, \frac{{\dot a}_\CD}{a_\CD}\average{\varrho}_{\CD}\;=\;0 \,.
\end{equation}
The equations \eref{eq:averagedraychaudhuri}, \eref{eq:averagedhamilton} and \eref{eq:energyconservation} constitute a system of macroscopic equations for the irrotational dust spacetime. Combining the three equations gives the already discussed integral constraint \eref{eq:integrability}.
This integral constraint suggests that backreaction and curvature are coupled, and that backreaction can induce changes to curvature. We may think collectively of backreaction and curvature as induced by backreaction as an effective large-scale \say{dark component} modifying the Friedmann equations of homogeneous cosmology. 

\section{Scaling solutions as a case study and a proof of concept}
\label{sec:scaling}
In this section we introduce a class of `scaling solutions' which satisfy the global field equations introduced in section \ref{sec:avfield}.
The equations \eref{eq:averagedraychaudhuri}, \eref{eq:averagedhamilton} and \eref{eq:integrability} form a set of three independent equations with four unknown macroscopic variables $a_\CD$, $\average{\varrho}_{\CD}$, $\average{\CR}_{\CD}$, and ${\CQ}_\CD$. A physically motivated closure condition is thus needed. 
Here we specify a class of closure conditions where backreaction obeys a powerlaw. We define an angular diameter distance model from a template metric construction. We then motivate a specific powerlaw model based on previous analyses, and analyze the ability of this model to be compatible with local measurements of the Hubble parameter when requiring convergence towards a slightly positively curved FLRW model consistent with Planck data \cite{Valentino}.

\subsection{A large-scale exact scaling solution}

Following a similar approach as in \cite{morphon2006,larena2009,roy:instability,Desgrange} we consider the following simple, but exact closure to the system of equations on the largest scales $\CD$:
\begin{equation}
\label{eq:scalingsol}
\average{\CR}_{\CD} \;=\;  {\mathcal W}_{\CD_0} \, a_{\CD}^{n} + 6k \, a_{\CD}^{-2}   \ ; \qquad {\CQ}_{\CD} \;=\;{\CQ}_{\CD_0} \, a_{\CD}^{n}\ , 
\end{equation}
which we denote \textit{scaling solutions}. 
While a generic irrotational dust spacetime obeys the set of equations discussed above, the equation \eref{eq:scalingsol} reduces the spacetimes under consideration to models where backreaction obeys a scaling law volume dependence on the largest scales identified with the domain $\CD$. The ansatz \eref{eq:scalingsol} is a simple extension of the FLRW powerlaw scaling of cosmological parameters.
The constant ${\mathcal W}_{\CD_0}$ is the backreaction-induced curvature ${\mathcal W}_\CD \equiv \average{\CR}_{\CD} - 6k \, a_{\CD}^{-2}$ evaluated at the present epoch, and $n$ is a scaling index determining the power law dependence with $a_{\CD}$. 
The `integration constant' term $6ka_{\CD}^{-2}$---where $k \equiv \Omega^{\tiny \textrm{FLRW} }_{k0} \, H^2_{\CD_0}$---is the Friedmannian component of the curvature that can be added to any solution satisfying \eref{eq:integrability} to obtain a new solution.
Plugging \eref{eq:scalingsol} into the integral constraint \eref{eq:integrability} provides the linear relation
\begin{equation}
\label{eq:backreactionpropcurv}
\average{\CR}_{\CD} \;=\;  6k \, a_{\CD}^{-2} \,-\, \frac{n\,+\,6}{n\,+\,2} \, {\mathcal Q}_{\CD} 
\end{equation}
between kinematical backreaction and the average spatial curvature, where the second term models the deviations from the Friedmannian behaviour that we below determine from perturbative considerations and observational data. We may rewrite the averaged energy constraint \eref{eq:hamiltonianC} as follows:
\begin{eqnarray}
\label{eq:OmegaX}
\hspace*{-1.8cm} \Omega_{m}^{\CD} +\, \Omega_\Lambda^\CD \,+\, \Omega_\CX^{\CD} \,+\, \Omega^{\tiny \textrm{FLRW} }_{k} \;=\, 1 \ ; \qquad \Omega_\CX^\CD \; \equiv \; \frac{4}{n+6} \Omega_{\mathcal W}^{\CD} \ ; \quad  \Omega^{\tiny \textrm{FLRW} }_{k} \equiv -\frac{k a^{-2}_{\mathcal{D}} }{ H^2_{\CD} } \ ,   
\end{eqnarray}
where $\Omega_{\mathcal W}^{\CD}  \equiv - {\mathcal W}_\CD /6/H_{\CD}^2$ is the 
backreaction-induced curvature parameter. 
The dimensionless cosmological parameter $\Omega_\CX^\CD$ can be seen as incorporating the collection of effects due to inhomogeneous structure (both the backreaction term itself ${\mathcal Q}_{\CD}$, but also the backreaction-induced curvature ${\mathcal W}_\CD$).  
Following \cite{Buchert2008}, we cast the equations \eref{eq:averagedraychaudhuri} and \eref{eq:averagedhamilton} into the form of the Friedmann equations by defining an additional \say{backreaction fluid component (indexed by $\textrm{b}$)} with effective density and pressure,
\begin{equation}
\label{eq:effectiverhop}
\varrho^{\textrm{b}}_{\CD} \equiv - \frac{{\mathcal Q}_{\CD}}{16 \pi G} - \frac{{\mathcal W}_{\CD}}{16 \pi G} \ ; \qquad   p^{\textrm{b}}_{\CD} \equiv - \frac{{\mathcal Q}_{\CD}}{16 \pi G} + \frac{{\mathcal W}_{\CD}}{48 \pi G} \ , 
\end{equation}
and with an effective dark energy equation of state
\begin{equation}
\label{eq:eqofstate}
w^{\textrm{eff}}_{\CD} \equiv \frac{p^{\textrm{b}}_{\CD}}{\varrho^{\textrm{b}}_{\CD} } = \frac{{\mathcal Q}_{\CD} - \frac{1}{3} {\mathcal W}_{\CD} }{ {\mathcal Q}_{\CD} + {\mathcal W}_{\CD} } = \frac{1 + \frac{1}{3} \frac{n\,+\,6}{n\,+\,2} }{1 - \frac{n\,+\,6}{n\,+\,2}} \ ; \qquad n \ne -2 \ ,
\end{equation}
where the scaling solution relation \eref{eq:backreactionpropcurv} has been used in the last equality. 
The effective dark energy equation of state is thus constant in the scaling solution case. 
Unlike for fundamental fields, \eref{eq:effectiverhop} and \eref{eq:eqofstate} are not \textit{a priori} constrained from energy conditions.
When $n = 0$, meaning constant backreaction, we have $w^{\textrm{eff}}_{\CD} = -1$ which mimics a scale-dependent cosmological constant; the above excluded index $n = -2$ mimics the evolution of a scale-dependent constant curvature model.

\subsection{Template metric, distances and structure-emergent curvature evolution}

Light propagation in inhomogeneous spacetimes is a highly non-trivial topic. While promising formalism have been proposed to study average properties of congruences of light and null-cones of observers \cite{Uzun,Mikolaj,lightconeav1,lightconeav2}, much remains to be understood for building consistent models for average photon propagation and observations in inhomogeneous spacetimes. 
It is of particular interest how volume-averaged variables defined over spatial sections relate to measurements of typical observers.

For spacetimes with a notion of statistical spatial homogeneity and isotropy, where structure is sufficiently slowly evolving such that the relevant timescales of structure formation are much larger than timescales over which null rays propagate over length scales of approximate statistical homogeneity, we expect null rays to probe spatial averages  \cite{lightpropndust,lightpropngeneral} and to have redshift which is given by the inverse volume scale factor.  
Thus, spatially averaged variables should provide information about light propagation in a broad class of spacetimes relevant for cosmology.

We employ the following template metric \cite{larena2009,Desgrange}---as motivated by Ricci flow smoothing of Riemannian hypersurfaces \cite{ricciflow}---to convert cosmological parameters into predictions for angular diameter distance for observations on the largest scales,
\begin{equation}
\label{eq:metric}
^4g^\CD  \equiv  - \dd t^2 \,+ L^2_{\CD_0}  a^2_\CD  \pa{\, \frac{\dd r_\CD^2}{1 - \kappa_\CD(t)  r_\CD^2} + r_\CD^2\,  \dd\Omega^2 \,}    \,,
\end{equation}
where $t$ labels the fundamental hypersurfaces of averaging, and $r_\CD$ is a dimensionless radial coordinate, which also has the interpretation as a comoving distance. 
\begin{equation}
L_{\CD_0} =
\left\{ \begin{array}{@{\kern2.5pt}ll}
    \hfill   \sqrt{ \left|  \Omega_\CR^{\CD_0} \right|    } H^{-1}_{\CD_0}  , &   \Omega_\CR^{\CD_0} \neq 0 \ , \\
      H^{-1}_{\CD_0}  , & \Omega_\CR^{\CD_0} = 0 \ , 
\end{array}\right.
\end{equation}
is the spatial curvature scale for curved models and the Hubble horizon in the spatially flat case. The dimensionless scale factor is set equal to unity at the present epoch $a_{\CD_0} = a_\CD(t_0) = 1$.
$\dd \Omega^2 \,\equiv\,  (\dd \theta^2 + \sin(\theta)^2 \, \dd \phi^2)$ is the angular element on the unit sphere, 
and $\kappa_\CD$ is a dimensionless spatial constant-curvature function 
\begin{equation} \label{eq:defkappa}
\kappa_\CD (t)\;\equiv 
\left\{ \begin{array}{@{\kern2.5pt}ll}
    \hfill  \,\frac{\average{\CR}_{\CD}(t)}{|{\average{\CR}_{\CD_0}}|} \, a^2_\CD(t) \ ,  &  \Omega_\CR^{\CD_0} \neq 0 \ , \\
      0 \ , & \Omega_\CR^{\CD_0} = 0 \ . 
\end{array}\right.
\end{equation}
The metric \eref{eq:metric} reduces to a spatial FLRW template metric on each spatial slice with scalar curvature equal to $\average{\CR}_{\CD}(t)/6$ on each spatial hypersurface of constant proper time $t=const.$, but the union of hypersurfaces does not in general correspond to a single four-dimensional FLRW metric. 
This dynamical curvature feature for the template metric reflects the lack of a `Newtonian' conservation law for the average three-Ricci scalar discussed in section \ref{sec:curvature}. 
We note that there are other possible extrapolations of the FLRW metric yielding the same spatial FLRW metric on each of the $t=const$ hypersurfaces, but which are associated with a different union of the surfaces into a four-metric. We employ the form of the metric \eref{eq:metric} in this analysis, keeping in mind the limited space of FLRW extrapolations investigated, \textit{c.f.} \cite{Koksbang}.
For investigations of the application of a spatially flat template metric in an interesting statistically homogeneous test case within the Buchert and Green \& Wald schemes, see \cite{SikoraGlod}, \textit{c.f.} \cite{CliftonSussman}.
The template metric is \textit{a priori} not a solution to the Einstein field equations, though see \cite{stichel} and references therein for investigations of \eref{eq:metric} as an exact solution to the Einstein field equations (with similar features for the curvature function that are found in this paper such as an initial positive curvature and a change of sign to negative curvature).
 
Let us consider a universe model with a statistically homogeneous and isotropic matter distribution which is slowly evolving compared to the time it takes for light to cross a homogeneity scale. The redshift associated with typical observers and emitters comoving with the slices of statistical homogeneity and isotropy and separated by distances larger than an approximate homogeneity scale is then well-approximated as \cite{lightpropndust,lightpropngeneral} $1+z = 1/a_\CD$. 
This identification of redshift is different from that used in \cite{larena2009,Desgrange} where the redshift function was calculated from the geodesic equation for null rays propagating on the `template metric background' \eref{eq:metric}. Such a phenomenological procedure is at odds for long-time evolution with the more rigorous calculations from local 
spacetime dynamics in \cite{lightpropndust,lightpropngeneral}, and we thus employ the approximative result $1+z = 1/a_\CD$ in the following analysis.\footnote{The two methods give comparable results in the low-redshift Universe, with $\sim 0.3 \%$ differences at the mean redshift of the Joint Lightcurve Analysis (JLA) supernova sample $z\sim 0.3$, but we here wish to span the whole cosmic epoch since decoupling. We skip the index $\CD$ at the redshift for notational ease.}
Note in this context that `Ricci-dominated metrics' (like the FLRW model that only features a Ricci curvature component) is at odds with the reality of light propagation in the sense that light predominantly `sees' the Weyl tensor (and it does so exclusively in the case of propagation through voids).

We assume that the angular diameter distance is well-described by that of the template metric, 
\begin{equation}
\label{eq:dA}
{\mathrm d}_A(z_\CD) \;=\;L_{\CD_0}  \, a_\CD(z_\CD) \, r_\CD(z_\CD) \, , 
\end{equation}
with $r_\CD(z_\CD)$ given by the radial null lines in \eref{eq:metric},
\begin{eqnarray}
\label{eq:dericd}
\frac{\dd r_\CD}{\dd a_{\CD}}\,= \,-  \frac{L_{\CD_0} H_{\CD_0}}{a_\CD^2}  \sqrt{\frac{1 \,-\, \kappa_{\CD}(a_{\CD}) \, r_\CD^2(a_{\CD})}{\Omega_{m}^{\CD_0} \, a_{\CD}^{-3}  \, + \Omega^{\tiny \textrm{FLRW} }_{k0} a_{\CD}^{-2} \, +\, \Omega_{\CX}^{\CD_0} \, a_{\CD}^{n} }} \ ,
\end{eqnarray}
with $r_\CD \,(a_{\CD} = 1) \, \equiv \, 0$.
With the scaling closure \eref{eq:scalingsol}, the macroscopic variables $a_\CD$, $\average{\varrho}_{\CD}$, $\average{\CR}_{\CD}$, ${\CQ}_\CD$ and the corresponding template metric are fully determined by the initial conditions. Assuming $\Lambda = 0$, the four parameters $\Omega_{m}^{\CD_0}$, $\Omega^{\tiny \textrm{FLRW} }_{k0}$, $H_{\CD_0}$, and $n$ uniquely determine the scaling solution. 

It is useful to consider the following curvature function
\begin{equation}
\label{eq:totalR}
 \frac{\average{\CR}_{\CD}(t) a^2_\CD(t) }{6 H^2_{\CD_0} } \, = \,    \left|  \Omega_{\CR}^{\CD_0}  \right|   \, \kappa_\CD (t) 
 =\,- \,    \Omega_{\mathcal W}^{\CD_0} a^{n+2}_\CD(t) - \Omega^{\tiny \textrm{FLRW} }_{k0}  \, , 
\end{equation}
where we have used the definitions given in \eref{eq:omega}, \eref{eq:OmegaX}, and \eref{eq:defkappa}.
The function \eref{eq:totalR} can be seen as an effective FLRW `present-epoch curvature parameter' for each hypersurface $t=const.$, and might be derived from the generic curvature statistic \cite{Clarkson} for models where an angular diameter distance and a Hubble parameter can be formulated as functions of redshift,
\begin{equation}
\label{eq:k}
k_H \,\equiv\, \frac{1}{D^2}\left(1 - \left(\frac{\dd D}{\dd z}\frac{H}{H_0}\right)^2 \right)\  , 
\end{equation}
where $D$ is related to the angular diameter distance $d_A$ by $D = H_0/c \, (1+z)d_A$, and where $H$ is the Hubble parameter of the model. We have omitted the label $\CD$ in the expression \eref{eq:k} for ease of notation, and any scale-dependence of $k_H$ remains implicit.
From the expression for the FLRW comoving distance $D = 1/ \sqrt{\Omega_{k0}} \, \sinh (  \sqrt{\Omega_{k0}} \, \int_{0}^{z} dz' \frac{H_0}{H(z')} )$---where $\Omega_{k0}$ is the FLRW cosmological curvature parameter evaluated at the present epoch---it follows that $k_H = - \Omega_{k0}$ in the FLRW class of metrics by identity.\footnote{This identity is purely geometrical, and robust to tuneable features within the FLRW class of metrics such as matter content, dark energy equation of state, and modifications of the Einstein field equations.}
In generic models the expression \eref{eq:k} need not coincide with a curvature parameter entering in an energy constraint equation, and it will in general fail to be a constant in redshift.  
Using the expressions for the angular diameter distance of the scaling solutions \eref{eq:dA}, \eref{eq:dericd}, we find that 
\begin{equation}
\label{eq:kscaling}
k_H \,=\,  \, \left|  \Omega_{\CR}^{\CD_0}  \right|  \, \kappa_\CD (t)    \, = \,  -  \Omega_{\mathcal W}^{\CD_0} a^{n+2}_\CD(t) - \Omega^{\tiny \textrm{FLRW} }_{k0}  \, , 
\end{equation}
which is equal to the curvature function in \eref{eq:totalR}. 
Thus, if one were able to determine the right-hand side of equation \eref{eq:k} model-independently\footnote{Model-dependent determinations of \eref{eq:k} should be treated with care as the assumptions might \textit{a priori} impose specific model curvature behaviour which need not correspond to that of the underlying spacetime.} at different redshifts, we would expect the outcome \eref{eq:kscaling} in the case of the scaling solution with scaling index $n$ being an accurate phenomenological model for describing the largest scales of the Universe.
For $n>-2$ we expect convergence to constant FLRW-type curvature at high redshifts, whereas at low redshifts it is expected to be dominated by curvature induced by structure formation. 

\subsection{Asymptotic positive FLRW curvature at the last scattering epoch}

We are interested in investigating whether we might be able to account for the apparent curvature discrepancy between Planck power spectra and low redshift datasets together with the FLRW Hubble parameter discrepancy within this model of backreaction. 
For the purpose of constraining the investigated class of scaling solutions, we fix the scaling index $n$ by its theoretical prediction obtained in \cite{Lagrangian2013}, \textit{cf.} \cite[sect.7.3]{Buchert2011}, in a Lagrangian perturbative framework around an Einstein--de Sitter background, where the leading-order backreaction was found to obey the scaling law ${\mathcal Q}_\CD \propto a_{\rm EdS}^{-1}$ corresponding to $n=-1$.\footnote{The scaling solution ${\mathcal Q}_\CD \propto a_{\mathcal{D}}^{-1}$ corrects the background-dependent scaling by following the domain scale factor $a_{\mathcal D}$ rather than $a_{\rm EdS}$, accounting for the volume difference in a curved space compared to that of a flat space.} 
See also \cite{NanLi} where backreaction in second-order standard perturbation theory in a dust universe model was also found to yield the scaling solution form \eref{eq:scalingsol} with $n=-1$. The scaling index $n=-1$ corresponds to an effective dark energy of state \eref{eq:eqofstate} of $w^{\textrm{eff}}_{\CD} = -2/3$ and is thus in between a cosmological constant scenario ($w^{\textrm{eff}}_{\CD} = -1$) and a scenario dominated by curvature ($w^{\textrm{eff}}_{\CD} = -1/3$).

Interestingly, the theoretical prediction $n=-1$ is supported by the fit of the scaling solutions with $\Lambda = 0$ to the Joint Lightcurve Analysis (JLA) dataset \cite{JLA2014} of type Ia supernovae,\footnote{In practice the luminosity distance--redshift relation of the scaling solutions is obtained by applying Etherington's reciprocity theorem.} keeping the scaling index as a free parameter \cite{Desgrange}. Fixing the scaling index to its theoretical prediction 
$n=-1$, the `$1\sigma$' confidence bounds on the matter cosmological parameter was constrained to be \cite{Desgrange} 
\begin{linenomath}
$$\Omega_{m}^{\CD_0} = 0.25^{+0.04}_{-0.04} \ ,$$ \end{linenomath} 
with a quality of fit comparable to that of the $\Lambda$CDM model. 
We note that this result was derived assuming $\Omega^{\tiny \textrm{FLRW} }_{k0} = 0$. 

However, for sufficiently small values of $\Omega^{\tiny \textrm{FLRW} }_{k0}$, the `energy budget' \eref{eq:OmegaX} at low redshifts is dominated by $\Omega_{m}^{\CD}$ and $\Omega_\CX^{\CD}$. The model $\Omega_{m}^{\CD_0} = 0.25 - \Omega^{\tiny \textrm{FLRW} }_{k0}/2$, $ \left| \Omega^{\tiny \textrm{FLRW} }_{k0}  \right|  \lesssim 0.05$\footnote{Where we compensate for the introduction of a non-zero value of $\Omega^{\tiny \textrm{FLRW} }_{k0}$ by modifying $\Omega_{m}^{\CD_0}$ and $\Omega_\CX^{\CD_0}$ by the same amount.} produces angular diameter distances of $\lesssim 0.5\%$ differences to the $\Omega_{m}^{\CD_0} = 0.25$, $\Omega^{\tiny \textrm{FLRW} }_{k0} = 0$ model for the redshift range $z \lesssim 1.3$ of the JLA sample. 
We thus conclude that we might safely use the modified best-fit model, 
\begin{linenomath}$$
\Omega_{m}^{\CD_0} = 0.25 - \Omega^{\tiny \textrm{FLRW} }_{k0}/2 \, ; \,  \left| \Omega^{\tiny \textrm{FLRW} }_{k0}  \right|\lesssim 0.05 \ ,
$$\end{linenomath} as an approximation for the purposes of this paper. 

Let us now consider the dark energy-free scaling solution model with $\Omega^{\tiny \textrm{FLRW} }_{k0} = - 0.04$---which converges (already at moderately high redshifts) toward an FLRW universe model with $\Omega_{k0} = - 0.04$ at early times---consistent with the best-fit value of the curvature parameter of the Planck power spectra \cite{Planck2018,Valentino}. Note that we might transform the Planck inferred FLRW curvature parameter $\Omega_{k0}$ to the corresponding value $\Omega_{k}(z^*)$ evaluated at the last scattering surface of central redshift $z^*$, and match the scaling solution to the FLRW model at this epoch by requiring $\Omega^{\tiny \textrm{FLRW} }_{k}(z^*) = \Omega_{k}(z^*)$. This matching procedure gives slightly different results than a simple matching at the present epoch $\Omega^{\tiny \textrm{FLRW} }_{k0} = \Omega_{k0}$ due to the different evolution of the Hubble parameter as a function of redshift in the models and the potential difference in redshift of the last scattering epoch. However, the difference in the evolution in Hubble parameter between the models and the model-independent constraint on redshift to the last scattering surface (see the below analysis) gives $\lesssim 0.002$ differences in these two determinations of $\Omega^{\tiny \textrm{FLRW} }_{k0}$, and our conclusions are robust to the exact choice of matching procedure.
    
From the best-fit supernovae result we thus construct the solution $\Omega_{m}^{\CD_0} = 0.25 - \Omega^{\tiny \textrm{FLRW} }_{k0}/2 = 0.27$, implying $\Omega_\CX^{\CD_0} = 0.77$ and $\Omega_{\mathcal W}^{\CD_0} = 5/4 \cdot 0.77$ from \eref{eq:OmegaX}. 
The value $\Omega_{m}^{\CD_0}  = 0.27$ is in good agreement with the recent model-independently determined matter density parameter from gas mass fraction measurements in galaxy clusters, supernovae observations and cosmic baryon abundance measurements from absorption systems at high redshifts \cite{Holanda}, $\Omega_{m}  = 0.285 \pm 0.013$, and with the $\Lambda$CDM inferred value from the Dark Energy Survey galaxy clustering and weak lensing \cite{DES}, $\Omega_{m}  = 0.267^{+0.030}_{-0.017}$.
These smaller values of $\Omega_{m}^{\CD_0}$ relative to that expected in $\Lambda$CDM are in line with what has been found in relativistic simulations done within a class of `silent universe models' \cite{Bolejko2017}, where the emergence of spatial curvature was found to cause $\Omega_{m}^{\CD_0}$ to be smaller, with $\sim 0.05$ relative to a $\Lambda$CDM model with the same initial conditions.

\begin{figure}[ht]
\centering
\includegraphics[width=0.8\textwidth,height=0.6\textwidth]{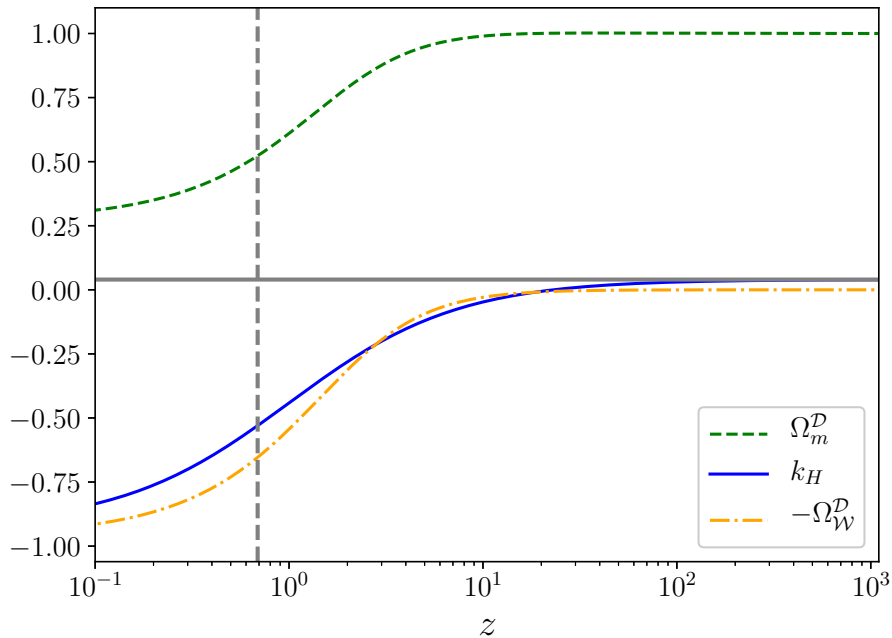}
\caption{The functions $k_H$, $- \Omega_{\mathcal W}^{\CD}$ and $\Omega_{m}^{\CD}$ as predicted by the scaling solution $n=-1$, $\Omega_{m}^{\CD_0}  = 0.27$, $\Omega^{\tiny \textrm{FLRW} }_{k0} = - 0.04$, and with zero cosmological constant. The horizontal line shows the $k_H$ line for an FLRW model with $\Omega_{k0} = -0.04$,  $k_H({\mathcal W}_{\CD_0} = 0)$, cf. eq. (20), which the scaling solution approaches asymptotically. For a FLRW universe model $k_H = - \Omega_{k0}$, where $\Omega_{k0}$ is the spatial curvature parameter evaluated at the present epoch. The dashed vertical line represents the redshift of transition from positive to negative volume deceleration.}
\label{fig:khscaling}
\end{figure}

The function $k_H$ for this model is shown in figure \ref{fig:khscaling}. 
At large redshifts we have asymptotic convergence towards $k_H = - \Omega^{\tiny \textrm{FLRW} }_{k0}$, whereas smaller redshifts are dominated by the negative induced average curvature due to structure formation. 

Figure~\ref{fig:khscaling} also displays the backreaction-induced curvature parameter $\Omega_{\mathcal W}^{\CD}$ and $\Omega_{m}^{\CD}$ of the scaling solution model.\footnote{Towards high redshifts the plot for $\Omega_{m}^{\CD}$ is not accurate due to the significant fraction of radiation at this epoch. The correction terms to the angular diameter distance to the epoch of last scattering from radiation is of order $\sim 0.5\%$ which is comparable or smaller than the errors on the estimates of the angular diameter distance used in this analysis. We thus neglect the contribution from radiation.} 
Backreaction-induced curvature reaches percent levels of the `energy budget' with $\Omega_{\mathcal W}^{\CD} \sim 0.01$ at $z=15$. Thus, our model reflects the conservative assumption of an almost FLRW universe model at early stages. The volume deceleration $q_{\CD} \equiv - {\ddot a}_\CD \, a_\CD /{\dot a}^2_\CD$ becomes negative for cosmic times corresponding to $z\sim 0.7$, as indicated by the dashed vertical line in figure~\ref{fig:khscaling}. The redshift of transition between volume deceleration and volume acceleration is comparable to that predicted by the current best-fit $\Lambda$CDM {model}.

The onset of backreaction-induced curvature is early as compared to what was found in numerical simulations within the `silent universe model' mentioned above \cite{Bolejko}, where $\Omega_{\mathcal W}^{\CD} \sim 0.01$ was reached at $z = 4$, which agrees with the general idea that backreaction effects might start to become significant when nonlinear structures start to form \cite{buchertrasanen}.
The early onset of backreaction in the scaling solution investigated is directly linked to the closure condition \eref{eq:scalingsol}, which only allows for backreaction and the associated average curvature to follow a powerlaw. Such a solution might not be adequate for extrapolation over large cosmological time intervals. 
We consider the scaling solution as a case study and proof of concept, keeping in mind that the extrapolation between redshifts probed by supernovae $z\lesssim 1$ and redshifts around the epoch of decoupling $z \sim1000$ is given by a large-scale leading-mode approximation. 

The magnitude of the curvature statistic $k_H$ at low redshifts shown in figure \ref{fig:khscaling} might seem to contradict the FLRW results in the literature, given the interpretation of $-k_H$ as an effective FLRW `present epoch' spatial curvature parameter at each $t=const.$ hypersurface. However, the tight combined constraints excluding a negatively curved universe model within the FLRW models, is not applicable to models which are not contained in the FLRW class of spacetimes. 

Interestingly, constraints on the FLRW curvature parameter $\Omega_{k0}$---independent of the theory of gravity on cosmological scales and matter content, but assuming the FLRW class of metrics---from supernovae and strong lensing probing redshifts $\lesssim 1.8$ hints at moderate negative curvature \cite{Collett} even though results are compatible with $\Omega_{k0} = 0$ at the level of one standard deviation. 
We also note that recent analysis \cite{NGS16_noacceleration} points to non-accelerating negatively curved FLRW universe models as good fits to data, especially when accounting for the dipole moment in the FLRW inferred acceleration \cite{Bernal,ColinMohayaee,Sarkar:SNshear}, suggesting that bulk flow caused by a regional under density can mimic the observed positive acceleration \cite{Tsagas}.
For the scaling solutions there is no local cosmological constant or energy component contributing to accelerating expansion---rather cosmological acceleration is an emergent large-scale effect caused by the rapidly expanding (almost) empty void regions gaining volume dominance in the late Universe. As already noted above, the transition from negative to positive global acceleration $\ddot{a}_\mathcal{D}$ occurs at a redshift $z\sim 0.7$ for the model investigated in this section, which is comparable to the onset of acceleration as estimated in the $\Lambda$CDM model.

The curvature statistic $k_H$ has been analyzed model-independently using the JLA sample, SDSS-III BOSS BAO measurements, and differential age measurements of galaxies \cite{MontanariRasanen}. In this analysis, tendencies for preferred negative $k_H$---corresponding to negative effective FLRW spatial curvature---was inferred (see their Fig.~6). Despite these tendencies, the analysis \cite{MontanariRasanen} showed consistency of the flat FLRW expectation $k_H = 0$ within $2\sigma$ confidence bounds.    

\subsection{Solving the Hubble discrepancy}

We now analyze the implications for the scaling solution Hubble parameter as inferred from the acoustic scale of the CMB. 
We use two different estimates of the redshift and angular diameter distance to the epoch decoupling. The results from Planck \cite{Planck2018} quoted in the first column of their table 2 gives $\{z^*_{\tiny \textrm{Planck} } = 1090.3 \pm 0.4$, $
d_{A \,\tiny \textrm{Planck} }(z^*_{\tiny \textrm{Planck} })/ \textrm{Mpc} = 12.72 \pm 0.05\}$. 
More model-independent constraints based on the allowance for a rescaling of the angular diameter distance to the epoch of decoupling in the Einstein--de Sitter model \cite{dAMI} gives the constraints $\{z^*_{\tiny \textrm{MI} } = 1094 \pm 1$, $d_{A \, \tiny \textrm{MI} }(z^*_{\tiny \textrm{MI} })/ \textrm{Mpc} = 12.7 \pm 0.2\}$. 
Requiring the angular diameter distance as parameterized by the scaling solution with $\Omega_{m}^{\CD_0}  = 0.27$ and $\Omega^{\tiny \textrm{FLRW} }_{k0} = - 0.04$ to coincide with the best-fit values of these empirical determinations of $\{z^*, \, d_{A}(z^*)\}$, we obtain:
\begin{eqnarray}
\label{eq:HfromDA}
\fl
\quad &H_{\CD_0} = 74.5 \; \textrm{km/s/Mpc} \qquad (\textrm{Planck: } z^* = 1090.3 \, ; \,  d_A(z^*)/ \textrm{Mpc} = 12.72 ) \ ; \nonumber \\
\fl
\quad &H_{\CD_0} = 74.2 \; \textrm{km/s/Mpc} \qquad  (\textrm{Model indep.: } z^* = 1094 \, ; \,  d_A(z^*)/ \textrm{Mpc} = 12.70 ) \ , 
\end{eqnarray}
with 1$\sigma$ error bars of order 5\% when taking into account the uncertainty on the matter cosmological parameter from the supernova analysis constraint. These best-fit results are in agreement with the low-redshift measurements of the expansion rate from cepheids and type Ia supernovae \cite{Riess2018,Riess2019}. Even though the results \cite{Riess2018,Riess2019} are derived in a $\Lambda$CDM 
model-dependent manner, the low-redshift value of the cepheids used to calibrate the Hubble parameter should make these measurements relative insensitive to the model cosmology and valid for comparison between models. 
The negative curvature component $\Omega_{\mathcal W}^{\CD}$, which falls off slowly with redshift due to the scaling index being $n=-1$, contributes to large distances to the CMB.

While the modification of the best-fit scaling solution from the JLA sample $\{\Omega_{m}^{\CD_0}  = 0.25$, $\Omega^{\tiny \textrm{FLRW} }_{k0} = 0 \}$ $\mapsto$ $\{\Omega_{m}^{\CD_0}  = 0.27$, $\Omega^{\tiny \textrm{FLRW} }_{k0} = - 0.04\}$ does not change the angular diameter distance--redshift relation significantly at the low redshifts probed by the JLA sample, it \emph{does} change the relation at high redshifts. 
For the model $\{\Omega_{m}^{\CD_0}  = 0.25$, $\Omega^{\tiny \textrm{FLRW} }_{k0} = 0 \}$ without a small positive FLRW curvature component the predictions \eref{eq:HfromDA} would have been $H_{\CD_0} = 82.2$ km/s/Mpc and $H_{\CD_0} = 81.9$ km/s/Mpc for the Planck and model-independent determination of $z^*,\, d_A(z^*)$, respectively. Thus, assuming the validity of the applied constraints for angular diameter distance and local Hubble parameter estimation in the context of the scaling solutions, a positive curvature component is not only allowed for in the scaling solutions but is necessary to fit the angular diameter distance to the recombination epoch.

\section{Discussion}
\label{sec:discussion}
A wide number of studies have been carried out in the field of inhomogeneous cosmology and the averaging problem, some of which directly address the Hubble parameter tension. 
In \cite{GiblinObservable} inhomogeneous corrections to the Hubble diagram were investigated in a periodic spacetime with a pressureless perfect fluid source with initial conditions as motivated from the $\Lambda$CDM framework and with observers and emitters placed on $t=$const.-hypersurfaces, employing coordinates in the BSSN (Baumgarte, Shapiro, Shibata, Nakamura) formalism \cite{SN,BS}. The mean Hubble diagram obtained was well in agreement with that expected from an FLRW model, with observer-dependent fluctuations in the angular diameter distance at redshifts $\sim 1$ of order $\lesssim 1\%$. 
These simulations were truncated at the level of the linear regime, with the observer hypersurface density contrasts being of order $\sim 2\%$, and it is perhaps not surprising that inhomogeneities have little effect both in terms of mean observables and variance between observers. 

A recent study has addressed inhomogeneity effects on the Hubble parameter \cite{AdamekHubble} in an N-body framework using a weak-field scheme based on smallness of field variables in the Poisson gauge \cite{gevolution,gevolution2}. 
Inhomogeneity effects were found to be insignificant for the mean luminosity distance of observers, while the variance in luminosity between observers was found to be of the order of a few percent for sources placed at redshifts $\sim 1$.
While the weak-field scheme employed in such analyses limits the class of relativistic spacetimes to those which obey certain smallness requirements on metric components and their derivatives (see \cite{gevolution2} for details), it offers a consistent framework to investigate deviations from homogeneity in a bigger class of relativistic models than those contained in the standard $\Lambda$CDM paradigm.\footnote{See also a recent study \cite{East} using an N-body simulation scheme with no restrictions of the metric components, but with simplified initial conditions.}
Apart from restrictions on the geometry made in these analyses, it will also make a difference that the observers are assumed to be comoving in the Poisson gauge---and not the comoving gauge coinciding with the frame of the matter constituting the spacetime. 

Within the same framework as that employed in \cite{AdamekHubble}, backreaction effects were found to be insignificant when the averaging operation was defined on $t=$const.-hypersurfaces in the Poisson gauge, large backreaction effects up to $\sim 15\%$ where found when averaging in the comoving gauge in this (coordinate-dependent) scheme \cite{AdamekSmoothing}.\footnote{In \cite{AdamekSmoothing} backreaction was quantified through average differences in expansion with respect to the \say{background} expansion rate, and their conclusions on the value of backreaction are thus not fully comparable to this paper which uses the conventional and background-independent definition of kinematical backreaction \cite{dust}, \textit{cf.} footnote~\ref{Qzero}.} 
It is a built-in feature of the covariant averaging scheme employed in our analysis that averaging is performed in the fluid 
frame,\footnote{Although see \cite{generalfluid} for cases where vorticity of the fluid is nonzero.} which reduces to the comoving gauge for spacetimes which are described in the weak-field limit around an FLRW background. Observers comoving in the Poisson gauge travel on extrinsic trajectories and therefore do not \textit{a priori} have physical significance, whereas the matter frame (the comoving gauge) has obvious physical significance as the frame with respect to which physical observers are at rest. 
For a discussion of the choice of frames and the validity of the linear extrapolation in FLRW background cosmology, see \cite{GiblinLimited,Clifton}. 
For a discussion of the choice of averaging frame for relativistic inhomogeneous spacetimes models, see \cite{BMR,foliationdependence}. 

A recent study \cite{MacphersonNumerical,MacphersonHubble} conveyed fully relativistic simulations using the Einstein Toolkit \cite{Etoolkit} for a universe model with closed spatial sections, and with dark matter modeled as a dust fluid\footnote{A small pressure component is added to the energy momentum tensor for numerical reasons, but the physical aim is to model a dust spacetime.} with initial conditions compatible with the $\Lambda$CDM scenario and observations of the cosmic microwave background. Here, insignificant backreaction effects were found in a gauge similar to the Poisson gauge on the largest scales of the simulations. In the same framework it was found that the variations in the Hubble parameter on spatial scales comparable to the supernovae catalogues used for determining the local Hubble expansion were not enough to account for the Hubble tension of $\Lambda$CDM cosmology. 

The Poisson gauge-inspired choice of averaging frame in these analyses was motivated by avoiding singularities generated from caustics in the codes, and is thus not adapted to any physical class of observers in the matter frame. Even though both the choice of foliation in which to perform the averaging and the definition of the backreaction variable differ from that employed in \cite{dust,perfectfluid,generalfluid} and the present analysis, the results are interesting and could potentially be straightforwardly adapted to a an approach where averaging is performed in (or close to) the matter frame. 

An example of relativistic fluid simulations with averaging performed in the matter frame is \cite{Bolejko}. 
Here general-relativistic simulations in the `silent universe model'---with relativistic ray-tracing based on the Sachs optical equations implemented---showed a backreaction-induced transition to negative curvature towards late cosmological epochs. 
While the silent universe model restricts the class of spacetimes considered to ones with no pressure gradients or energy flux, it offers a realistic background-free modeling of the late Universe on scales larger than regions dominated by relativistic speeds, heat flow, and gravitational wave and rotational degrees of freedom.
Within the same simulation, initial data consistent with Planck were shown to generate a Hubble parameter consistent with low-redshift measurements \cite{Riess2018,Riess2019} as a direct consequence of the emergent negative spatial curvature.
The overall conclusions of this analysis agree with the results of our analysis. 

It is interesting to ask whether the differences in outcome of the simulations discussed are indeed related to the choice of averaging frame and the frame in which the model observers are placed. We note that the choice of averaging is a physical choice (not a choice of coordinates) of how we define the macroscopic theory. 
Though nothing in principle prevents us from carrying out an analysis where the averaging frame is adapted to a coordinate system which is for instance convenient for numerical simulation, the resulting set of spatial hypersurfaces are not necessarily the ones where macroscopic variables defined from averaging have a clear physical interpretation. 
For instance, if the volume scale factor is to be interpreted in terms of inverse mean redshift---which is in practice assumed in many analyses employing averaging---then this imposes constraints on the choice of hypersurfaces which must likely be chosen to be close to the matter frame \cite{Rasanenreview}. 
The macroscopic average variables, such as the average restmass density are directly (and correctly) interpretable when defined in the fluid frame. In addition, any averaging operation employed must assure that the same collection of fluid elements is averaged in the course of time \cite{covariance,BMR}. When considering observers these should be placed in this same frame if they are to represent typical physical observers in the Universe. 

\section{Summary and Conclusion}
\label{sec:conclusion}
We have discussed the assumptions about spatial curvature which are inherent in the FLRW ansatz usually imposed in cosmological analysis. We have discussed how the situation differs in a generic average model subject to the laws of general relativity where curvature and structure in the matter distribution are dynamically coupled. Models with dynamical curvature are not \textit{a priori} excluded from any physical principle nor from any existing cosmological 
dataset, rather they are natural in a general-relativistic universe model with structure on a hierarchy of scales.  

As a case study of models allowing for spatial curvature we have considered a class of scaling solutions which are obtained by imposing constraints on the generic solutions of the averaged Einstein equations. 
We have shown that the best-fit scaling solution from the JLA sample of type Ia supernovae accounts for the Hubble parameter anomaly in FLRW cosmology when an asymptotic FLRW curvature parameter coinciding with that preferred from the Planck power spectra \cite{Planck2018,Valentino} is required in the model. 
The coincidence of the best-fit scaling solution obtained from type Ia supernovae data fitting the peak of the angular diameter distance with an asymptotic FLRW spatial curvature parameter of $\Omega_{k0} = -0.04$ indicates that 
general-relativistic dynamical spatial curvature models are natural candidates for accounting for the tensions between high- and low-redshift cosmological datasets in FLRW cosmology.      

In FLRW cosmology the preferred positive curvature of the Planck power spectra arises from an enhanced lensing 
signal---resulting in smoothing of high multipoles---relative to what would be expected within a flat $\Lambda$CDM model \cite{Planck2015,Planck2018}. 
It is beyond the scope of this paper to aim for quantifying lensing within the scaling solutions which would require quantification in a yet undeveloped perturbative framework (for the architecture of a possible framework see \cite{Roy2012}). Here we merely point out that large-scale dynamical curvature models---as exemplified by the scaling solutions---can account for positive curvature in the early Universe while being consistent with local expansion rate measurements as a result of dynamical, structure-emergent average curvature \cite{dust}. 

The reader may recall that our model simply assumed a large-scale leading-mode approximation for backreaction, which is a member of a generic realization of average properties of the $3+1$ Einstein equations, together with consistent and physically motivated metric and distance notions. As a proof of concept, this model 
respects (i) the generic coupling of geometry (curvature) to the sources, (ii) the generic non-conservation of the average scalar curvature, and (iii) it reflects the generic possibility of the change of sign of the averaged scalar curvature.
The result is a natural and consistent explanation of (i) dark energy, (ii) the coincidence problem, (iii) positive initial curvature, (iv) the small matter density cosmological parameter found in local probes of the matter density, (v) the large angular diameter distance to the CMB consistent with JLA supernova sample parameter constraints, and (vi) the local expansion rate measurements (removal of the `Hubble tension'). 

We believe that this model architecture needs convincing arguments to be rejected as a physically viable show-case,
on the basis of which the model ingredients can be improved in order to build a physical cosmology in the future.

\vspace{8pt}
{\bf Acknowledgements:}
{\small This work is part of a project that has received funding from the European Research Council (ERC) under the European Union's Horizon 2020 research and innovation programme (grant agreement ERC advanced grant 740021--ARTHUS, PI: TB). We wish to thank L\'eo Brunswic for discussions on topology, Jan Ostrowski for many valuable comments during this work, and Krzysztof Bolejko, Syksy R\"as\"anen, Peter Stichel and David Wiltshire for fruitful comments on the manuscript.
We thank the referees for their critical and constructive comments that helped us improve this paper.}


\end{document}